\documentclass{KapProc} 
\kluwerbib 
 
 \normallatexbib  
\begin{document} 
\newcommand{\hmp}{h^{-1}Mpc}       
\newcommand{\Ga}{\Gamma}      
\newcommand{\Om}{\Omega}      
\newcommand{\de}{\delta}      
\newcommand{\al}{\alpha}      
\newcommand{\si}{\sigma}      
\newcommand{\bx}{{\bf x}}      
\newcommand{\lam}{\lambda}      
\newcommand{\lan}{\langle}      
\newcommand{\ran}{\rangle}      
\newcommand{\La}{\Lambda}      
\newcommand{\bm}{\boldmath}      
\newcommand{\be}{\begin{equation}}      
\newcommand{\ee}{\end{equation}}      
\newcommand{\bea}{\begin{eqnarray}}      
\newcommand{\eea}{\end{eqnarray}}      
\newcommand{\ra}{\rightarrow}      
\newcommand{\bef}{\begin{figure}}      
\newcommand{\eef}{\end{figure}}      
\newcommand{\Mpc}{{\rm Mpc}}      
        
\newcommand{\veps}{\varepsilon}       
\def\spose#1{\hbox to 0pt{#1\hss}}       
\def\ltapprox{\mathrel{\spose{\lower 3pt\hbox{$\mathchar"218$}}       
 \raise 2.0pt\hbox{$\mathchar"13C$}}}       
\def\gtapprox{\mathrel{\spose{\lower 3pt\hbox{$\mathchar"218$}}       
 \raise 2.0pt\hbox{$\mathchar"13E$}}}       
\def\inapprox{\mathrel{\spose{\lower 3pt\hbox{$\mathchar"218$}}       
 \raise 2.0pt\hbox{$\mathchar"232$}}}       
\articletitle{Correlation and clustering } 
 
\articlesubtitle{Statistical properties of galaxy large scale structures}

\author{Francesco Sylos Labini} 
\affil{D\'ept.~de Physique Th\'eorique, Universit\'e de Gen\`eve\\ 
24, Quai E. Ansermet, CH-1211 Gen\`eve, Switzerland} 
\email{sylos@amorgos.unige.ch} 
 
\author{Andrea Gabrielli} 
\affil{INFM Sezione Roma 1\\ 
Dip. di Fisica, Universita' ``La Sapienza'',  
P.le A. Moro 2 I-00185 Roma, Italy. } 
\email{andrea@pil.phys.uniroma1.it}

\begin{abstract} 
In this lecture we  
clarify the basic difference between  the correlation 
properties for  systems characterized by small    
or large fluctuations. The concepts of correlation 
length, homogeneity scale, scale invariance and  
criticality are discussed as well. 
We relate these concepts to the interpretation of galaxy clsutering.
\end{abstract} 
 
 
\section{Introduction} 
 
The existence of large scale structures (LSS)     
and voids in the distribution of galaxies     
up to  several hundreds Megaparsecs is well known from  
twenty years     
\cite{huchra,tully}. 
The relationship of these structures     
with  the statistics of galaxy distribution     
is usually inferred by applying the      
standard statistical analysis as introduced     
and developed by Peebles and coworkers \cite{pee80}. 
Such an analysis {\it assumes} implicitly that      
the distribution is homogeneous at very small scale     
($\lambda_0 \approx 5 \div 10 \hmp$).  
Therefore the system is characterized as having small fluctuations      
about a finite average density.    
If the galaxy distribution had a fractal nature  
the situation would be completely different.      
In this case the average density in finite samples is not  
a well defined quantity: it is strongly sample-dependent  
going to zero in the limit of an infinite volume.       
In such a situation it is not meaningful to study     
fluctuations around the average density extracted from sample data.     
The statistical properties of the distribution should    
then be studied in a completely     
different framework than     
the standard one. We have been working     
on this problem since some time   
\cite{slmp98}   
by following the original ideas of Pietronero \cite{pie87}.     
The result is that galaxy structures are     
indeed fractal up to tens of Megaparsecs   
\cite{joyce99a}.     
Whether   a crossover     
to homogeneity at a certain scale  $\lambda_0$,      
occurs or not (corresponding to the absence     
of voids of typical scale larger than $\lambda_0$)       
is still  a matter of debate  \cite{rees99}.     
At present, the problem is basically that the available redshift surveys     
do not sample scales larger than $50 \div 100 \hmp$     
in a wide portion of the sky and in a complete way.     
 
In this lecture 
we try to clarify some simple and basic concepts like 
the proper definition of correlation length, homogeneity scale,  
average density and scale invariance. We point out that a  
correct defintion and intepretation of the above  
concepts is necessary in order to understand phenomenologically 
the statistical properties of galaxy structures and  
to define the correct theoretical questions one would like 
to answer for.

\section{Distribution with Small Fluctuations}

Consider a {\em statistically}    
homogeneous and isotropic particle density     
$n(\vec{r})$ with or without correlations with a well defined average  
value $n_0$. Let  
\be 
\label{corr1b} 
n(\vec{r}) = \sum_i \delta(\vec{r} - \vec{r}_i) 
\ee 
be the number density of points in the system (the index $i$ runs  
over all the points) and let us suppose to have an infinite system. 
Statistical homogeneity and isotropy refer to the fact that   
any $n$-point statistical property of the system   
is a function only on the scalar relative distances between these $n$ points.  
The existence of a well defined average density means that   
\be 
\lim_{R\rightarrow\infty}  
\frac{1}{\|C(R)\|}\int_{C(R)}d^3r\, n(\vec{r})=n_0>0 
\ee 
(where $\|C(R)\|\equiv 4\pi R^3/3$ is the volume of the sphere   
$C(R)$) independently of the origin of coordinates.   
The scale $\lambda_0$, such that  the {\it one point} average density 
is well-defined, i.e.  
\be 
\left|\int_{C(R)}d^3r\, n(\vec{r})/\|C(R)\|-n_0\right|<
n_0 \;\; \mbox{for} \;\;  
r>\lambda_0 \;, 
\ee 
is  called {\em homogeneity scale}.  
If $n(\vec{r})$ is extracted from a density ensemble,   
$n_0$ is considered the same for each realization, i.e. it is   
a self-averaging quantity.    
 
Let $\left<F \right>$ be the ensemble average of a    
quantity $F$ related to  $n(\vec{r})$.  
If only one realization of $n(\vec{r})$ is available,    
$\left<F \right>$ can be evaluated as an average over    
{\em all} the different points (occupied or not) of the space  
taken as origin    
of the coordinates.  
The quantity    
\[\left<n(\vec{r_1})n(\vec{r_2})...n(\vec{r_l})\right>dV_1dV_2...dV_l\]    
gives the average probability of finding $l$ particles placed     
in the infinitesimal volumes $dV_1,dV_2,...,dV_l$ respectively    
around $\vec{r_1}, \vec{r_2},...,\vec{r_l}$.    
For this reason    
$\left<n(\vec{r_1})n(\vec{r_2})...n(\vec{r_l})\right>$ is called     
{\em complete} $l$-point correlation function.    
Obviously $\left<n(\vec{r})\right>=n_0$, and in a single sample such that   
$V^{1/3}\gg \lambda_0$, it can be   
estimated  by  
\be 
n_V=N/V    
\label{ave} 
\ee 
where $N$ is the total number of particle in volume $V$.    
  
Let us analyze the auto-correlation properties of such a system.    
Due to the hypothesis of statistical homogeneity and isotropy,     
$\left<n(\vec{r_1})n(\vec{r_2})\right>$ depends only on    
$r_{12}=|\vec{r_1}-\vec{r_2}|$.    
Moreover,    
$\left<n(\vec{r_1})n(\vec{r_2})n(\vec{r_3})\right>$ is only a function of     
$r_{12}=| \vec{r_1}-\vec{r_2}|$, $r_{23}=| \vec{r_2}-\vec{r_3}|$ and    
$r_{13}=|\vec{r_1}-\vec{r_3}|$.    
The {\em reduced} two-point and three correlation functions $\xi(r)$   
and $\zeta(r_{12},r_{23},r_{13})$ are respectively defined by:  
\bea  
&&\left<n(\vec{r_1})n(\vec{r_2})\right>=n_0^2\left[1+\xi(r_{12})\right]   
\label{03a}\\  
&&\left<n(\vec{r_1})n(\vec{r_2})n(\vec{r_3})\right>=    
n_0^3   
\left[1+\xi(r_{12})+\xi(r_{23})+   
\xi(r_{13})+\zeta(r_{12},r_{23},r_{13})\right]\,.\nonumber   
\label{xi} 
\eea  
The reduced two-point correlation function $\xi(r)$ defined in the 
previous equation is a useful tool to describe  
the correlation properties of small fluctuations  
with respect to the average. However  we stress  
again that in order to perform  
a statistical analysis following Eqs.\ref{xi}, 
the one-point average density should be a well  
defined quantity and this must be carefully 
tested in any given sample (see below). 
We define          
\be 
\label{sigma1} 
\sigma^2(R) \equiv \frac{\langle \Delta N(R) ^2\rangle } 
{\langle N(R)\rangle } 
\ee 
to be the mean square fluctuation normalized to the 
average density.  
From the definition of $\lambda_0$, we have 
\be 
\label{sigma2} 
\sigma^2(\lambda_0) \simeq 1 
\ee 
and $\sigma^2(R) \ll 1$ for $r \gtapprox \lambda_0$. 
Note that $\sigma^2(R)$ is again related to the {\it one-point} 
property of the distribution. We stress that the defintion 
of the homogeneity scale via Eq.\ref{sigma2} can be 
misleading in the case where the average density is not 
a well-defined concept (see next section). Indeed, in such 
a case the quantity $\langle N(R)\rangle$ at the denominator 
{\it is not} given by $\langle N(R)\rangle= n_0 \times \|C(R)\|\sim R^3$: 
the scaling exponent is indeed different from the Euclidean 
dimension of the space $d=3$. 
 
In order to analyze observations    
from an occupied point it is necessary    
to define another     
kind of average:    
the {\em conditional} average $\left<F \right>_p$ 
which characterizes the {\it two-point} properties of the system.    
This is defined as an ensemble average with the    
condition that the origin of coordinates    
is an occupied point.   
When only one realization of $n(\vec{r})$    
is available,  $\left<F \right>_p$  
can be evaluated averaging the quantity $F$     
over all the occupied points taken as origin of coordinates.      
The quantity  
\be 
\left<n(\vec{r_1})n(\vec{r_2})...n(\vec{r_l})\right>_pdV_1 dV_2...dV_l 
\ee 
is  the average probability of finding $l$ particles placed     
in the infinitesimal volumes $dV_1,dV_2,...,dV_l$    
respectively around $\vec{r_1}, \vec{r_2},    
...,\vec{r_l}$ with the condition that    
the origin of coordinates is an occupied point.    
We call $\left<n(\vec{r_1})n(\vec{r_2})...n(\vec{r_l})\right>_p$    
conditional $l$-point density.    
Applying the rules of conditional probability \cite{feller} one has:         
\bea  
\label{03c}  
&&\Gamma(r) \equiv \langle n(\vec{r}) \rangle_p   
=\frac{\langle n(\vec{0})n(\vec{r}) \rangle}{n_0}\\  
&&\langle n(\vec{r_1})n(\vec{r_2}) \rangle_p =   
\frac{\langle n(\vec{0})n(\vec{r_1})n(\vec{r_2}) \rangle}{n_0}\,.  
\nonumber  
\eea  
where $\Gamma(r)$ is called the conditional  
average density \cite{pie87}.  
   
However, in general, the following convention is assumed  
in the definition of the conditional densities:    
the particle at the origin does not observe itself.    
Therefore $\langle n(\vec{r}) \rangle_p$ is defined only for     
$r>0$, and $\langle n(\vec{r_1})n(\vec{r_2}) \rangle_p$   
for $r_1, r_2>0$.     
In the following we use this convention as     
corresponding to the  
experimental data in galaxy catalogs.       
 
We have defined above the homogeneity scale by means of 
the one-point properties of the distribution. Here we 
may define it in another wa  by looking at the two-point properties: 
If the presence of an object at the point $\vec{r}_1$  
influences the probability of finding another object  
at $\vec{r}_2$,  
these two points are correlated. Hence  there is a correlation 
at the scale distance $r$ if  
\be 
\label{corr1} 
G(r) = \langle n(\vec{0})n(\vec{r})\rangle   \ne \langle n\rangle  ^2  \,. 
\ee 
On the other hand, there is no correlation if 
\be 
\label{corr2} 
G(r) = \langle n\rangle  ^2>0\,. 
\ee 
Therefore the proper definition of  $\lambda_0$, the {\it homogeneity scale}, 
is  the length scale beyond which 
$G(r)$ 
or equivalently $\Gamma(r)$ become nearly constant with scale and show 
a well-defined flattening. If this scale is smaller than the sample size then 
one may study for instance the behaviour of $\sigma^2(R)$ with scale 
(Eq.\ref{sigma1}) in the sample. 
 
The length-scale $\lambda_0$ represents the typical dimension 
of the voids in the system. 
On the other hand there is another length scale which is very important 
for the characterization of point spatial distributions: the {\em correlation  
length} $r_c$. 
The length $r_c$   
separates scales at which density fluctuations are correlated (i.e.  
probabilistically related) to scales where they are uncorrelated. 
It can be defined only if a crossover towards homogeneity is  
shown by the system, i.e. if  
$\lambda_0$ exists \cite{perezmercader}. 
In other words $r_c$ defines the   organization 
in geometrical structures of the fluctuations  
with respect to the average density. Clearly 
$r_c > \lambda_0$: 
only if the average density can be defined  
one may study the correlation length of the fluctuations 
around it. 
Note that $r_c$ {\it is not} 
related to the absolute amplitude of fluctuations, 
but to their probabilistic correlation. 
In the case in which $\lambda_0$ is finite  
and then $\langle n \rangle >0$, in order  
to study the correlations properties of the fluctuations around the 
average and then the behaviour of $r_c$, we can  
study  
the reduced two-point correlation function $\xi(r)$ defined in  
Eq.\ref{xi}. 
 
The {\it correlation length} can be defined through the 
scaling behavior of $\xi(r)$ with scale. 
There are many definitions of $r_c$, but in any case, in order to have  
$r_c$ finite, $\xi(r)$ must decay enoughly fast to zero with scale. 
For instance, if  
\be 
\label{and2} 
|\xi(r)| \rightarrow \exp(-r/r_c)\;\;\mbox{for}\;r \rightarrow \infty\,, 
\ee 
this means that for $r\gg r_c$ the system is  
structureless and density fluctuations are weakly correlated. 
The definition of the  
correlation length $r_c$ by Eq.\ref{and2} is 
equivalent to the one given by  
\cite{perezmercader}. 
 
\subsection{Some Examples}  
 
Let us consider some simple examples. 
The first one is a Poisson distribution for  
which there are no correlation between different points. 
In such a situation  
\cite{gsl2001} 
the average density is well defined, 
and  
\be 
\xi(r_{12}) = \delta(\vec{r_1}-\vec{r_2})/ n_0  \,. 
\ee  
Analogously, one can obtain the three point correlation functions:    
\be 
\zeta(r_{12},r_{23},r_{13})=  
\delta(\vec{r_1}-\vec{r_2})\delta(\vec{r_2}-\vec{r_3})/n_0^2\;. 
\ee 
The two previous relations    
say only that there is no correlation     
between different points. That is, the reduced    
correlation functions $\xi$    
and $\zeta$ have only the so called {\it ``diagonal''} part.    
This diagonal part is present in the reduced correlation functions of any     
statistically homogeneous and isotropic distribution with correlations.    
For instance \cite{saslaw} $\xi(r)$ in general can be written as     
$\xi(r)=\delta(\vec{r})/n_0+h(r)\,$,    
where $h(r)$ is the non-diagonal part which is meaningful only for $r>0$.   
Consequently, we obtain for the purely Poisson case (remember that  
conditional densities are defined only for points out of the origin):     
\bea  
\label{con-poi}  
&&\langle n(\vec{r})\rangle_p=n_0 \\   
&&\langle n(\vec{r_1})n(\vec{r_2})\rangle_p=n_0^2   
\left[1+\delta(\vec{r_1}-\vec{r_2})/n_0\right]\;.  
\nonumber  
\eea 
 
The second example is a distribution which is homogeneous 
but with a finite correlation lenght $r_c$.  
In such a situation $\Gamma(r)$ has a well-defined flattening 
and one may study the properties of $\xi(r)$. 
The correlation length $r_c$  is usually    
defined as the scale beyond which $\xi(r)$    
is exponentially damped. 
It measures up to which    
distance density fluctuations     
density are correlated.  
Note that while $\lambda_0$    
refers to an one-point property of the system    
(the average density), $r_c$ refers to a two-points    
property (the density-density correlation)    
\cite{perezmercader,gsld00}. 
In such a situation $\xi(r)$ is in general 
represented by  
\be 
\xi(r) = A  \exp(-r/r_c) 
\ee 
where $A$ is a prefactor which basically depends on  
the homoegeneity scale $\lambda_0$. We remind that 
$\lambda_0$ gives the scale beyond which $\sigma^2(R) \ll 1$  
(Eq.\ref{sigma1}), and not the scale beyond which  
density fluctuations are not correlated anymore.  
This means that the typical dimension of voids in  
the system is not larger than $\lambda_0$, but  
one may find structures of density fluctuations of 
size up to $r_c \ge \lambda_0$. 
 
Finally, let us now consider a mixed case in which the system    
is homogeneous (i.e. $\lambda_0$ is finite),    
but it has long-range power-law correlations. 
This means that fluctuations around the average, independently on 
their amplitude,  
are correlated at all scales, i.e. one finds 
structures of all scales. However, we stress again 
these are structures of fluctuations with respect to a mean 
which is well defined.     
This last event is in general described by the divergence of    
the  correlation length $r_c$. 
Therefore let us consider a system in which    
$\left<n(\vec{r})\right>=n_0>0$    
and $\xi(r)=[\delta(\vec{r})]/n_0+h(r)$, with    
\be |h(r)|\sim r^{-\gamma} \;\;\;\;\mbox{for}\;r\gg \lambda_0\,,    
\label{sc-in} 
\ee 
and  $0<\gamma\le 3$. For $\gamma >3$, despite the power law behavior,  
$\xi(r)$ is integrable for large $r$, and  
depending on the studied statistical quantity of the point distribution, 
we can consider the system as having 
a finite $r_c$ (i.e. behaving like an exponentially damped $\xi(r)$) or not. 
Eq.~\ref{sc-in} characterizes the presence of scale-invariant structures 
of fluctuations with long-range correlations, which, in Statistical Physics  
is also called ``critical'' \cite{huang}. 
 
\section{Distribution with Large Fluctuations} 
 
A completely different case of point distribution with respect the  
homogeneous one with or without correlations is the fractal one. 
In the case  of a fractal distribution, the average 
density $\left<n\right>$ in the infinite system is zero, then  
$G(r)=0$ and $\lambda_0 = \infty$ 
and consequently $\xi(r)$ is not defined. 
For a fractal point distribution with dimension $D<3$     
the conditional one-point density $\left<n(\vec{r})\right>_p$    
(which is hereafter called $\Gamma(r)$) has the following behavior    
\cite{slmp98}      
\be 
\label{fra0} 
\left<n(\vec{r})\right>_p\equiv \Gamma(r)=Br^{D-3}\;,  
\ee 
for enough large $r$. 
The intepretation of this behavior is the following. 
We may compute  
the average  mass-length relation     
from an occupied point which gives the average numebr of points  
in a spherical volume of radius $R$ centered on an occupied point: 
this gives       
\begin{equation}      
\label{f1}      
\langle  N(R) \rangle_p=      
(4\pi B)/D  \times R^D \; ,      
\end{equation}      
The constant $B$ is directly related to the lower cut-off        
of the distribution: it gives the mean number of galaxies
in a sphere of radius $1 \hmp$.       
Eq.\ref{f1} implies that     
the average density in a sphere of radius $R$ around    
an occupied point scales as $1/R^{3-D}$.     
Hence it depends on the sample size $R$, the    
fractal is asymptotically empty and thus     
$\lambda_0\rightarrow\infty$.    
We have two limiting cases for the fractal dimension: 
(1) $D=0$ means that there is a  
finite number of points well localized far from the 
boundary of the sample 
(2) $D=3$ the distribution has a well defined positive average density, 
i.e. the conditional average density does not depend on scale anymore. 
Given the metric interpretation of the fractal dimension, 
it is simple to show that $0 \le D \le 3$. 
Obviously, in the case $D=3$ 
for which $\lambda_0$ is finite $\Gamma(r)$ provides the 
same information of $G(r)$, i.e. it  
characterizes the crossover to homogeneity. 
 
A very important point 
is represented by the kind of information  
about the correlation properties 
of the infinite system which  can be extracted from the analysis 
of a finite sample of it. 
In \cite{pie87}  
it is demonstrated that, in the hypothesis of statistical homogeneity and 
isotropy, even in the super-correlated  
case of a fractal the estimate of $\Gamma(r)$  
extracted from the finite sample of size $R_s$, is  
not dependent on the sample size $R_s$,  
providing a good approximation of  
that of the whole system. Clearly this is true a part from statistical  
fluctuations \cite{gsl2001} due to the finiteness of the sample. 
In general 
the $\Gamma(r)$ extracted from a sample can be written in the following way: 
\be 
\label{and0} 
\Gamma(r) = \frac{1}{N} \sum_{i=1}^{N} \frac{1}{4 \pi r^2 \Delta r} 
\int_{r}^{r+\Delta r} n(\vec{r}_i+\vec{r}')d^{3}r', 
\ee 
where $N$ is the number of points in the sample, 
$n(\vec{r}_i+\vec{r}')$ is the number of 
points in the volume element $d^3r'$ around the point  
$\vec{r}_i+\vec{r}'$ and $\Delta r$ 
is the thickness of the shell at distance $r$ from the point at $\vec{r}_i$. 
Note that the case of a sample of a homogeneous point distribution of size 
$V\ll \lambda_0^3$, must be studied in the same framework of the fractal case. 
 
\section{Problems of the standard analysis} 
 
In the fractal case ($V^{1/3}\ll\lambda_0$),  
the sample estimate of the  
homogeneity scale, through the value of $r$ for which the  
sample-dependent correlation function  
$\xi(r)$ (given by Eq.\ref{and0b}) is equal to $1$, is meaningless. 
This estimate is the  so-called {\it``correlation length''}  
$r_0$ \cite{pee80} in the standard approach of  
cosmology. As we discuss below, 
 $r_0$ has nothing to share with the {\it true} correlation length $r_c$. 
 Let us see why $r_0$ is unphysical in the case $V^{1/3}\ll\lambda_0$. 
The length $r_0$  \cite{pee80}  is 
 defined by the relation $\xi(r_{0})= 1$, 
where $\xi(r)$ is given operatively by 
\be 
\label{and0b} 
\xi(r)=\frac{\Gamma(r)}{n_{V}} -1 \; . 
\ee 
where $n_V$ is given by Eq.\ref{ave}. 
What does $r_0$ mean in this case ? 
The basic point in the present discussion \cite{pie87}, 
is that the mean density of the sample, $n_V$, 
used in the normalization of  $\xi(r)$, is not an intrinsic quantity  
of the system, 
but it is a function of the finite size $R_s$ of the sample. 
 
Indeed, from Eq.\ref{fra0},    
the expression of the $\:\xi(r)$ of the sample in the case of 
fractal distributions is \cite{pie87} 
\be 
\label{xi3} 
\xi(r) =  
\frac{D}{3} \left( \frac{r}{R_s} \right)^{D-3} -1 \; . 
\ee 
being $R_s$ the radius of the assumed spherical sample of volume $V$. 
From Eq.\ref{xi3} it follows that $\:r_0$ (defined as $\:\xi(r_{0}) = 1$) 
is a linear function of the sample size $\:R_{s}$ 
\be 
\label{xi4} 
r_{0} =\left(\frac{D}{6}\right)^{\frac{1}{3-D}}R_{s} 
\ee 
and hence it is a spurious quantity without  physical meaning but it is 
simply related to the sample's finite size.  
In other words, this is due to the fact that $n_V$ in the fractal case is in 
any finite sample a {\em bad} 
estimate of the asymptotic density which is zero in this case 
 
We note that the amplitude of $\Gamma(r)$ (Eq.\ref{fra0}) 
is related to the lower 
cut-off of the fractal, while the amplitude of $\xi(r)$ 
is related to the upper cut-off (sample size $R_s$) of the distribution.  
This crucial difference has never been appreciated  appropriately. 
 
Finally, we stress that in the standard analysis of galaxy catalogs 
the fractal dimension is estimated by fitting $\xi(r)$ with a power law,  
which instead, as one can see from 
 Eq.\ref{xi3}, is power law only for  $r \ll r_0$ (or $\xi \gg 1$). 
For distances around and beyond $r_0$ there is a clear deviation 
from the power law behavior due to the definition of $\xi(r)$. 
Again this deviation is due to the finite size of the observational  
sample and does not correspond to any real change 
in the correlation properties. It is easy to see that if  
one estimates the exponent  
at distances $r \ltapprox r_0$, one  
systematically obtains a higher value  
of the correlation exponent due to the break  
in $\xi(r)$ in a log-log plot.

\section{Discussion and Conclusion} 
 
From an operative point of view, 
having a finite sample of points (e.g. galaxy catalogs), 
the first analysis to be done is the determination of  
$\Gamma(r)$ of the sample itself. Such a measurement 
is necessary to distinguish 
between the two cases: 
(1) a crossover towards 
 homogeneity in the sample with a flattening of $\Gamma(r)$, 
and hence an estimate of $\lambda_0<R_s$ and $\langle n \rangle$; 
(2) a continuation of the fractal behavior. 
Obviously only in the case (1), it is  
physically meaningful to introduce an estimate 
of the correlation function $\xi(r)$  
(Eq.\ref{and0b}), and extract from it the length scale 
 $r_0$ ($\xi(r_0) = 1$) 
to estimate the intrinsic homogeneity scale $\lambda_0$. 
In this case, the functional behavior of $\xi(r)$ with distance  
gives instead information on the correlation length of 
the density fluctuations. Note that there are always 
subtle finite size effects which perturb the behvaiour of $\xi(r)$
for $r \sim V{1/3}$, and which must be properly taken into account.
These same arguments apply to the estimation of the power spectrum
of the density fluctuations, which is just the fourier conjugate
of the correlation function \cite{slmp98}. 
The application of these concepts to the case of real  
galaxy data can be found in \cite{slmp98,joyce99a,gsl2001}. 
 
\begin{acknowledgments} 
We thank Y.V. Baryshev, R. Durrer, J.P.Eckmann, P.G. Ferreira, M. Joyce,     
M. Montuori and L. Pietronero for useful discussions.     
This work is partially supported by the       
 EC TMR Network  "Fractal structures and  self-organization"        
\mbox{ERBFMRXCT980183} and by the Swiss NSF.       
\end{acknowledgments}

\end{document}